# Comparison between historic nuclear explosion yield formulas

J.P. Lestone[a] and M. D. Rosen[b]

[a]Computational Physics Division, XCP-7, Los Alamos National Laboratory, Los Alamos, NM 87545
[b]Livermore National Laboratory, Livermore, CA 94550

January 20th, 2021

**Abstract:** During the Manhattan project a simple formula was developed by Bethe and Feynman in 1943 to estimate the yield of a fission-only nuclear explosion of a uniformly-dense bare-sphere of supercritical fissile material. We have not found any evidence that Bethe and Feynman knew of the first yield formula obtained by Frisch and Peierls contained within their famous March 1940 memorandum. Similarly, we have not found any technical documents that compare the Bethe-Feynman formula to the earlier works of Frisch and Peierls, or Serber. After adjusting for differences in the labeling of critical radii, we find that earlier formulas only differ by a scaling factor.

**Introduction**

From the beginning of the atomic age many famous authors, including Fermi, Oppenheimer, Compton, and Pryce, thought and speculated about a simple formula for the yield of a fission-only explosion[1] containing relatively easy-to-obtain inputs. Some of the earliest well documented works on a yield formula of fission-only nuclear explosions were those of Frisch and Peierls,[2] Serber,[3] Bethe and Feynman, and Pryce and Dirac. The earlier two of these have long been available in the open literature.

The Bethe-Feynman formula was developed at Los Alamos during the Manhattan project in 1943, to provide a straightforward way to estimate the yield of a fission-only nuclear explosion. Weapons codes containing details of hydrodynamics and neutronics, and the corresponding computing facilities were in their infancy[4,5], and there was an advantage to being able to estimate the yield without undertaking a time-consuming calculation of the full details of the explosion process. Here we present the Bethe-Feynman formula and describe its relationship to the other WWII era formulas. The reader should be aware that the Frisch-Peierls memorandum was published in the appendix of Serber's "The Los Alamos Primer"[3] with transcription errors. This appendix is not a part of Serber's work, and was added to the 1992 book as an editorial discussion. The Frisch and Peierls memorandum is available, error-free, in Gowing's 1964 classic on early atomic-energy research by the British.[6]

Strangely, the relationship between the first yield formula by Frisch and Peierls (1940), and that of Bethe and Feynman (1943), appears not to have been discussed before. Bethe and Feynman, and Serber, do not reference the Frisch-Peierls memorandum. It is possible that they did not know of it, as the initial communication of this memorandum to the USA, like the UK MAUD report[7], would have had a small readership. But having said that, from 1944 on, both Peierls and Frisch were in Los Alamos, and one might have expected them to have discussed any similarities between their formula and the corresponding Bethe and Feynman results. Three of the four actually shared the same Theoretical Division building at Los Alamos. However, we have not found any evidence that the relevant authors made such a comparison. However, a letter from Peierls to Oppenheimer, March 15th 1944, in our National Security Research Center archives[8] does illuminate some limited technical interactions in 1944 on theoretical calculations of efficiency between Bethe in the US, and Peierls and Dirac in Britain.

**Bethe and Feynman**

The yield formula by Bethe and Feynman was first derived on or before October 1943. Their formula for the yield from the nuclear explosion of a uniformly-dense bare-sphere of supercritical fissile material is

$$Y = f_{BF} M R_0^2 \alpha_0^2 \delta, \qquad (1)$$

where $M$ is the mass of the sphere; the initial radius $R_0$, and the initial time required for the neutron population to multiply by a factor $e$, $\tau_0 = 1/\alpha_0$ are both at the time of explosion. The dimensionless factor $\delta = R_2/R_0 - 1$, where $R_2$ ($> R_0$) is the radius of the exploding (expanding) system as it reverts back into a subcritical state. This is often referred to as the radius at second critical. First critical occurs during the device assembly phase. The reader should be aware that the above formula is the end result of a time dependent approach developed by Bethe and Feynman, namely the formula is the yield at infinite time, at the end of Bethe and Feynman's dynamical description.

To enable a comparison of the Bethe-Feynman formula to the earlier work of Frisch and Peierls, we consider a scenario like Frisch and Peierls' original suggestion in which a supercritical spherical assembly of $^{235}$U is made by bringing two subcritical nominal density items together, after which it explodes and expands to the point at which its reduced density causes it to cross "second critical" and become subcritical (an example is given in Table I). Notice that $R_2 > R_0 > R_C$; $\rho_2 < \rho_0$; and $\rho_0 = \rho_C$. In the limit of low yield, the mass and isotopics of the initial assembly and the expanded assembly at second critical are the same. With the additional simplifying assumption of a homogeneous explosion[3], we can then write

$$\rho_0 R_0^3 = \rho_2 R_2^3. \quad (2)$$

Given that the critical mass, $M_C$, at a nominal density of $\rho_C = \rho_0$, and the exploding sphere of mass $M_0 = M_2$ at second critical, must both have the same $\rho r$, we can write

$$\rho_2 R_2 = \rho_0 R_C = 158.4 \text{ g/cm}^2. \quad (3)$$

Here we use a $^{235}$U critical mass $M_C = 46.6$ kg [9] at a nominal density of 18.9 g/cc, and thus with a $R_C = 8.4$ cm. Substituting $\rho_2$ from Eq. (2) into Eq. (3) we get

$$\rho_0 \frac{R_0^3}{R_2^3} R_2 = \rho_0 R_C \Rightarrow \frac{R_0^2}{R_2^2} = \frac{R_C}{R_0}, \quad (4)$$

and thus (pg. 39[3])

$$\frac{R_2}{R_0} = \sqrt{\frac{R_0}{R_C}}. \quad (5)$$

The reason for the detailed discussion to obtain this equation is that with it we can rewrite Eq. (1) as

$$Y = f_{BF} M R_0^2 \alpha_0^2 \left(\sqrt{R_0/R_c} - 1\right). \quad (6)$$

Table I. The assembling of an 80.0 kg sphere of $^{235}$U at a nominal initial density of $\rho_0$=18.9 g/cc. (for illustrative purposes).

| Description | mass (kg) | density (g/cc) | radius (cm) |
|---|---|---|---|
| initial assembly | $M_0$=80.0 | $\rho_0$=18.9 | $R_0$=10.0 |
| 2$^{nd}$ critical | $M_2$=80.0 | $\rho_2$=14.4 | $R_2$=11.0 |
| critical mass | $M_C$=46.6 | $\rho_C$=18.9 | $R_C$= 8.4 |

**Frisch and Peierls**

The first documented example of an atomic-weapon yield formula is in the famous Frisch and Peierls March 1940 memorandum[2] with "… the energy liberated before the mass expands so much that the reaction is interrupted"

$$E \sim f_{FP} M (R_0/\tau_0)^2 \left(\sqrt{R_0/R_c} - 1\right). \quad (7)$$

Excluding the difference in scaling factor, the Frisch and Peierls' formula is the same as the Bethe-Feynman formula, as given in Eq. (6). Frisch and Peierls estimated the scaling factor in Eq. (7) to be $f_{FP} \sim 0.2$ "by a rough calculation …" Unfortunately the logic and assumptions behind their rough calculation appear to have been lost. The previous inability to notice the similarities between these two formulas is likely due to the many copies of the Frisch and Peierls memorandum where the square root in the yield formula incorrectly extends over the "−1", and possibly some confusion of the definition of the terms related to the critical radii $R_2$ and $R_C$.

The Frisch and Peierls memorandum contains the first "super-bomb" yield estimate of a potentially working bomb at 10 kt ($4\times10^{20}$ ergs). This initial yield estimate is within a factor of approximately two of both WWII devices. Many copies of the Frisch and Peierls memorandum have an expected energy release of $4\times10^{22}$ ergs, i.e. a factor of one hundred larger than the value given above. This huge yield is inconsistent with the input values quoted in the Frisch and Peierls memorandum, and is another commonly propagated transcription error.

**Serber**

The yield formula by Serber is documented in "The Los Alamos Primer".[3] This was the first technical document issued by the Los Alamos Laboratory (L.A. 1) and summarizes five lectures given by Serber in April 1943. On pg. 42[3] Serber states "the efficiency is of the order of"

$$f \sim (1/6)(v'^2/\varepsilon \tau^2) R_c^2 \Delta \quad (8)$$

where $\varepsilon$ is "the energy release for complete conversion"; $v'/\tau = \alpha_0$; and $\Delta = 2(R_2 - R_0)/R_C$. A simple rearrangement of Eq. (8) in the case of a slightly critical system ($\Delta \ll 1$) gives the nuclear yield (energy release)

$$Y \sim f_S M R_0^2 \alpha_0^2 (R_2/R_0 - 1), \quad (9)$$

with $f_S = 1/3$ (subscript "S" for Serber). On pg. 43 of "The Los Alamos Primer" Serber remarks how Oppenheimer was displeased with the cavalier way in which Serber had treated the hydrodynamic expansion of the fuel assembly as it explodes. It may be speculated then, that the contributions of Bethe and Feynman were an outgrowth of Oppenheimer's desire to see a more detailed description of the energy production of the exploding assembly as it expanded. However, excluding the difference in scaling factor, the Serber formula, in the limit of slightly critical systems, is the same as the Bethe-Feynman formula as given in Eq. (1), and with the inclusion of Eq. (5) the same as the Frisch and Peierls results given in Eq. (7).

**Pryce and Dirac**

Independent of Bethe and Feynman, a very similar formalism was developed by Pryce and Dirac in the UK. The Bethe-Feynman formalism used an ansatz of how alpha drops in time as the assembly's radius grows. Pryce and Dirac did not use this ansatz, but instead used more detailed neutronic considerations to obtain their relationship. There are other practical differences between the two approaches, but they are beyond the scope of this paper. However, Pryce and Dirac obtained formulas similar to those discussed above, namely the appearance of the same dependence on $M$, $R_0$, $\alpha_0$, and $\delta$, but with scaling factors, $f_{PD}$, that can depend on other device details.

**Conclusions**

Based on the early, and apparently independent, works of Frisch and Peierls,[2] Serber,[3] Bethe and Feynman, and Pryce and Dirac, it appears that it is relatively easy to determine that the yield of fission-only nuclear explosions scale with the fissile mass, $M$; the fissile material radius squared, $R_0^2$, and the neutronic reactivity squared, $\alpha_0^2$, both at the time of explosion; and the relative fissile material radius change needed for the system to revert back into a subcritical state, $\delta$. The agreement between these early studies on the main terms of relevance to a fission-only explosion appears to be not commonly known. The ease by which these relationships can be obtained partially justifies why the formulas of Frisch and Peierls, and Serber have long been available in the open literature.

The central role played by Frisch and Peierls' 1940 memorandum, in starting the UK and US nuclear weapons efforts, is undisputed. However, the similarity of their yield formula with those of Serber, Bethe and Feynman, and Pryce and Dirac, is often lost due to commonly propagated transcription errors, and confusions about the definition of critical radii.

**Acknowledgment**

We thank M. B. Chadwick for suggesting this as important area of study, and for suggesting that the Frisch and Peierls yield formula might be very similar to the corresponding results of others. We thank P. Adsley for information regarding the work of Pryce and Dirac.

This work was supported by the US Department of Energy through the Los Alamos National Laboratory. Los Alamos National Laboratory is operated by Triad National Security, LLC, for the National Nuclear Security Administration of the US Department of Energy under Contract No. 89233218CNA000001.

[b] This work was performed under the auspices of the U. S. Department of Energy by Lawrence Livermore National Laboratory under Contract DE-AC52-07NA27344.